\documentclass{iaus}
\usepackage{graphicx}

\title[~~What makes a galaxy radio-loud?] 
{What makes a galaxy radio-loud?}

\author[R. A. Ortega-Minakata, J. P. Torres-Papaqui, H. Andernach, et al.]
{R. A. Ortega-Minakata$^1$, J. P. Torres-Papaqui$^1$, H. Andernach$^1$,\\
R. Coziol$^1$, J. M. Islas-Islas$^1$, I. Plauchu-Frayn$^2$, D. M. Neri-Larios$^1$\\
\and M. del C. Rojas-Granados$^3$}

\affiliation{$^1$Departamento de Astronom\'ia, Universidad de Guanajuato,
C.P. 36000, Guanajuato, Mexico\\email: {\tt rene@astro.ugto.mx}\\[\affilskip]
$^2$Instituto de Astrof\'isica de Andaluc\'ia (CSIC),
E-18008, Granada, Spain\\[\affilskip]
$^3$Divisi\'on de Ingenier\'ias, Universidad de Guanajuato,
C.P. 36885, Salamanca, Mexico
}

\pubyear{2011} 
\volume{284}  
\pagerange{1--12}
\jname{The Spectral Energy Distribution of Galaxies}
\editors{R.J. Tuffs \&  C.C.Popescu, eds.}
\begin{document}

\maketitle

\begin{abstract}
We compare the Spectral Energy Distribution (SED) of radio-loud
and radio-quiet AGNs in three different samples observed with SDSS:
radio-loud AGNs (RLAGNs), Low Luminosity AGNs (LLAGNs) and AGNs in
isolated galaxies (IG-AGNs). All these galaxies have similar optical
spectral characteristics.
The median SED
of the RLAGNs is consistent with the characteristic SED of quasars,
while that of the LLAGNs and IG-AGNs are consistent with the SED of
LINERs, with a lower luminosity in the IG-AGNs than in the LLAGNs.
We infer the masses of the black holes (BHs) from the bulge masses.
These increase from the IG-AGNs to the LLAGNs and are highest for the RLAGNs.
All these AGNs show
accretion rates near or slightly below 10\% of the Eddington limit, the
differences in luminosity being solely due to different BH
masses.
Our results suggests there are two types of AGNs, radio quiet
and radio loud, differing only by the mass of their bulges or BHs.

\end{abstract}

\firstsection 
\section{Introduction}

The Spectral Energy Distribution (SED) of galaxies is a
tool that could allow to make a physical and possibly evolutionary
connection between galaxies showing different levels of AGN
activity. In this study, we compare the SEDs of radio-loud and radio-quiet
AGNs in three different samples with the typical SEDs of QSOs
and quasars. Our samples are composed of SDSS galaxies which are
Radio Loud (RLAGNs) with extended radio structures
(\cite[Lin \etal\ 2010]{lin10}), Low Luminosity AGNs (LLAGNs),
which turn out to reside mostly
in groups and clusters
(\cite[Torres-Papaqui \etal\ 2011]{tor11}), and AGNs in
isolated galaxies (IG-AGNs,
\cite[Coziol \etal\ 2011]{coz11}). The SEDs are based on radio flux
densities from NVSS (1.4\, GHz), FIR magnitudes from
IRAS (100, 60, 25, 12 $\mu$m) and optical fluxes from SDSS (5100
\AA). Except for a few detections with Chandra ($\sim$ 5 keV) in
IG-AGNs (5 of 25 radio-detected IG-AGNs), no X-rays were found 
for the other AGNs. We used the stellar population synthesis code {\sc
Starlight} to subtract a template from which we deduce the
bulge mass of the galaxies.

\section{Sample Description and Results}

\begin{figure}[h]
\begin{center}
\includegraphics[width=0.98\linewidth]{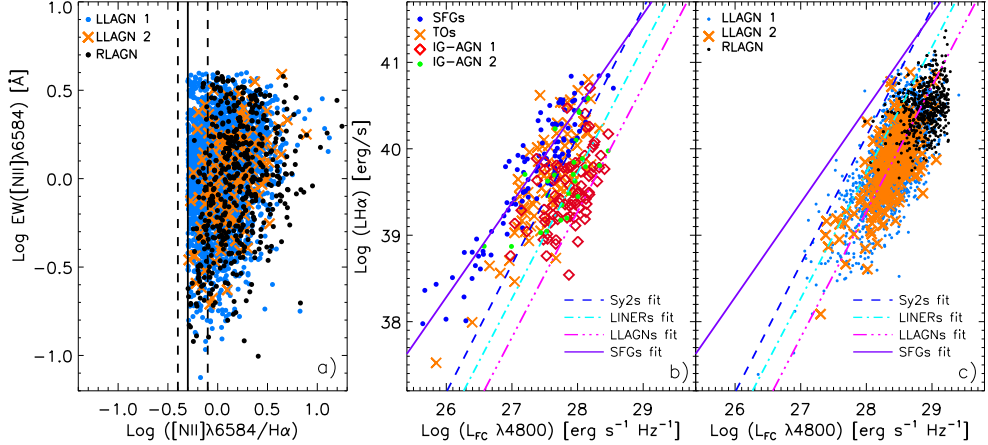}
 \caption{ a) NII diagnostic diagram used to identify the LLAGNs
and RLAGNs; b) and c) relations between the continuum luminosity at
4800 \AA\ and the H$\alpha$ luminosity for the AGNs in IG-AGNs (b) and
the LLAGNs (c). For comparison, in (b) and (c) we show power law fits for
Sy2, LINERs, LLAGNs, and Star Forming Galaxies (SFGs), based on a total of
$\sim$ 3$\times$10$^5$ SDSS galaxies
(\cite[Torres-Papaqui \etal\ 2011]{tor11}).}
   \label{fig1}
\end{center}
\end{figure}

The RLAGNs and LLAGNs in our sample are mostly early-type galaxies
in groups or clusters. Very few (142 of 4197) LLAGNs are detected in radio
(NVSS at 1.4\, GHz).
Despite the difference in radio emission the RLAGNs and LLAGNs show
similar spectra in the optical: both are narrow-line emission
galaxies, frequently with some emission lines missing
([OIII]$\lambda$5007 and/or H$\beta$). The NII diagnostic diagram,
Fig.\,\ref{fig1}a, was used to determine the nature of their activity
(\cite[Torres-Papaqui \etal\ 2011]{tor11}). Galaxies with
$log$~[NII]/H$\alpha > -0.3$ and $log$~EW[NII]~$< 0.6$
are classified as LLAGNs.

The IG-AGNs are mostly spiral galaxies with intense narrow emission
lines. A standard diagnostic diagram was used to identify 104 AGNs among the 292 IGs.
Only 25 of these were detected in NVSS.
In Fig.\,\ref{fig1}b we show that both the radio-undetected (IG-AGN~1) and
radio-detected (IG-AGN~2) IG-AGNs are mostly LINERs. In Fig.\,\ref{fig1}c 
we show that the RLAGNs are generally more luminous than both the
radio-undetected (LLAGN~1) and radio-detected (LLAGN~2) LLAGNs.

\begin{figure}[h]
\begin{center}
\begin{minipage}{0.69\linewidth}
\includegraphics[width=\textwidth]{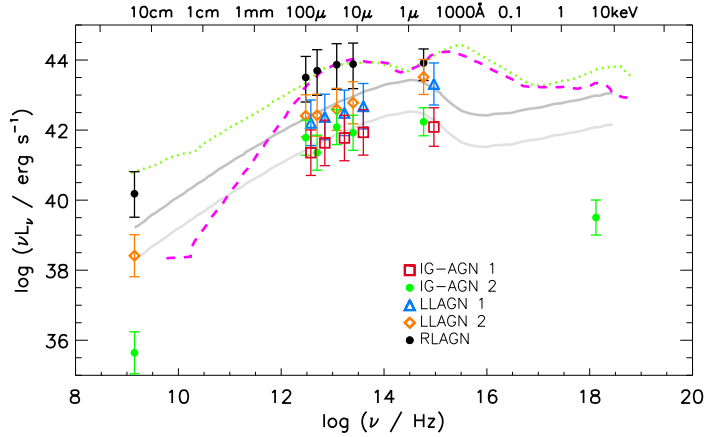}
\end{minipage}\hfill
\begin{minipage}{0.29\linewidth}
 \caption{
The SEDs of the different AGNs in our sample are compared with
characteristic SEDs for quasars (dotted curve), QSOs (dashed curve)
and LINERs (solid curves; one each for the LLAGNs and IG-AGNs).
The symbols with error bars represent the median and quartiles.
The data points for the IG-AGN~1 and LLAGN~1 samples were shifted
slightly to the blue for clarity sake. }
   \label{fig2}
\end{minipage}
\end{center}
\end{figure}

In Fig.\,\ref{fig2} we show the median SEDs of the AGNs in our
sample, as compared with characteristic SEDs for quasars, QSOs and LINERs
(\cite[Elvis \etal\ 1994]{elv94}; \cite[Younes \etal\ 2010]{you10}).
The SEDs of the quasars and QSOs were scaled down in
luminosity to fit the optical data for the RLAGNs. The LINER SED was
scaled up in luminosity to fit the optical and FIR data for the
LLAGNs and the IG-AGNs. The observed radio power is generally lower
than expected from the characteristic SEDs and models. The few X-ray
detections with Chandra of galaxies in the IG-AGN~2 sample also fall
much below expectation.

\begin{figure}[h]
\begin{center}
\includegraphics[width=0.98\linewidth]{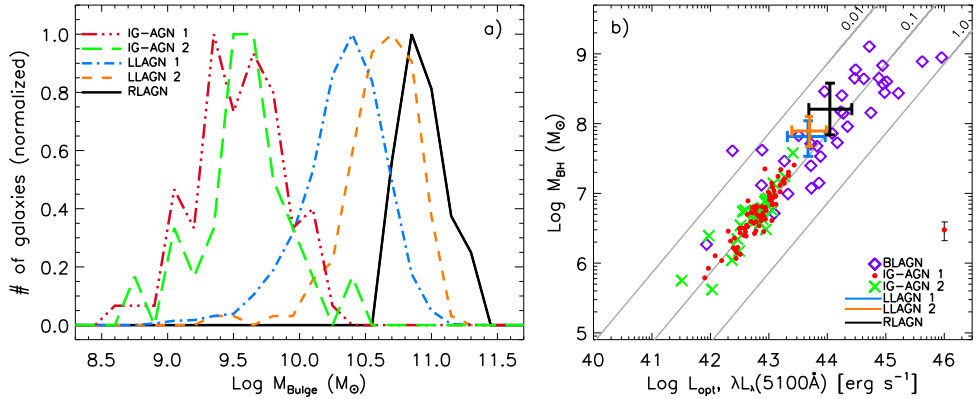}
 \caption{
a) Normalized histograms for the bulge masses; b) relation between
continuum luminosity at 5100\AA\ and black hole mass as deduced from
the mass of the bulge. The data for broad line AGNs (BLAGNs) are
from \cite{pet05}. }
   \label{fig3}
\end{center}
\end{figure}

In Fig.\,\ref{fig3}a we find that the mass of the bulge decreases in
the following order: RLAGNs, LLAGN~2, LLAGN~1,
IG-AGN~2 and IG-AGN~1. In each sample, the radio-loud galaxies always
have the more massive bulges.
In Fig.\,\ref{fig3}b we show that all the black holes (BHs) accrete at
rates near or slightly below 10\% of the Eddington limit. The
differences in luminosity seem solely due to different BH
masses, inferred from the relation by
\cite{har04}.

\section{Discussion and Conclusions}

The similarity of the optical spectral characteristics of RLAGNs and
LLAGNs suggests they have a common nature, implying they
are intrinsically two types of AGNs, radio loud and radio quiet.
The median SEDs of LLAGNs and IG-AGNs are consistent with scaled-up versions
of the characteristic SED of LINERs, while that of RLAGNs is consistent with
a scaled-down SED of quasars.
The difference in bulge mass between radio-loud and radio-quiet AGNs
suggests the central BH mass to be higher in radio-loud AGNs.
The fact that the IG-AGNs, which have formed and evolved in
isolation, have smaller BH masses, suggests that the
environment of a galaxy may not only determine its morphology but
also its AGN type.\\[-02mm]

J.P. T-P acknowledges support grants by PROMEP (103.5-10-4684), and
DAIP-UGto (65/11).
I. P-F acknowledges postdoctoral fellowship \#145727 by CONACyT, Mexico.

\end{document}